\documentclass[twocolumn,showpacs,amsmath,amssymb]{revtex4}
\usepackage{amsmath} %

\usepackage{graphicx}
\usepackage{amsfonts}
\usepackage{dcolumn}
\usepackage{bm}

\begin{document}

\title{Nucleation in scale-free networks}

\author{Hanshuang Chen$^1$}

\author{Chuansheng Shen$^{1,2}$}

\author{Zhonghuai Hou$^1$}\email{hzhlj@ustc.edu.cn}

\author{Houwen Xin$^1$}

\affiliation{$^1$Hefei National Laboratory for Physical Sciences at
 Microscales and Department of Chemical Physics, University of
 Science and Technology of China, Hefei, 230026, China \\ $^2$Department of Physics, Anqing Teachers College,
Anqing 246011, China}

\date{\today}

\begin{abstract}

We have studied nucleation dynamics of the Ising model in scale-free
networks with degree distribution $P(k)\sim k^{-\gamma}$ by using
forward flux sampling method, focusing on how the network topology
would influence the nucleation rate and pathway. For homogeneous
nucleation, the new phase clusters grow from those nodes with
smaller degree, while the cluster sizes follow a power-law
distribution. Interestingly, we find that the nucleation rate
$R_{Hom}$ decays exponentially  with the network size $N$, and
accordingly the critical nucleus size increases linearly with $N$,
implying that homogeneous nucleation is not relevant in the
thermodynamic limit. These observations are robust to the change of
$\gamma$ and also present in random networks. In addition, we have
also studied the dynamics of heterogeneous nucleation, wherein $w$
impurities are initially added, either to randomly selected nodes or
to targeted ones with largest degrees. We find that targeted
impurities can enhance the nucleation rate $R_{Het}$ much more
sharply than random ones. Moreover, $\ln (R_{Het}/R_{Hom})$ scales
as $w^{\gamma-2/\gamma-1}$ and $w$ for targeted and random
impurities, respectively. A simple mean field analysis is also
present to qualitatively illustrate above simulation results.

\end{abstract}
 \maketitle

\section{Introduction} \label{sec1}

Complex networks describe not only the pattern discovered
ubiquitously in real world, but also provide a unified theoretical
framework to understand the inherent complexity in nature
\cite{RMP02000047, PRP06000175, PRP08000093,SIR03000167}. Many real
networks, as diverse as ranging from social networks to biological
networks to communication networks, have been found to be scale-free
\cite{SCI99000509}, i.e., their degree distributions follow a
power-law, $P(k)\sim k^{-\gamma}$. A central topic in this field has
been how the network topology would influence the dynamics taking
place on it. Very recently, critical phenomena in scale-free
networks (SFNs) have attracted considerable research interest
\cite{RMP08001275}, such as order-disorder transitions
\cite{PHA02000260, PLA02000166, PRE02016104, EPB02000191},
percolation \cite{PRL00004626, PRL00005468, PRL02208701,
PRE02036113}, epidemic spreading \cite{PRL01003200}, synchronization
\cite{PRL03014101, PRL07034101}, self-organized criticality
\cite{PRL03148701, PRE002065102}, nonequilibrium pattern formation
\cite{Nat10000544}, and so on. These studies have revealed that
network heterogeneity, characterized by diverse degree
distributions, makes the critical behaviors of SFNs quite different
from those on regular lattices. However, most previous studies
focused on evaluating the onset of phase transition in different
network models. There is little attention paid to the
dynamics/kinetics of phase transition, such as nucleation and phase
separation in complex networks.

Nucleation is a fluctuation-driven process that initiates the decay
of a metastable state into a more stable one \cite{Kashchiev2000}. A
first-order phase transition usually involves the nucleation and
growth of a new phase. Many important phenomena in nature, including
crystallization \cite{JCP97003634}, glass formation
\cite{PRE98005707}, and protein folding \cite{PNAS9510869}, etc.,
are associated with nucleation. Despite its apparent importance,
many aspects of nucleation process are still unclear and deserve
more investigations. The Ising model, which is a paradigm for many
phenomena in statistical physics, has been widely used to study the
nucleation process. Despite its simplicity, the Ising model has made
important contributions to the understanding of nucleation phenomena
in equilibrium systems and is likely to yield important insights
also for nonequilibrium systems. In two-dimensional lattices, for
instance, shear can enhance the nucleation rate and at an
intermediate shear rate the nucleate rate peaks \cite{JCP08134704},
a single impurity may considerably enhance the nucleation rate
\cite{JPC06004985}, and the existence of a pore may lead to
two-stage nucleation and the overall nucleation rate can reach a
maximum level at an intermediate pore size. Nucleation pathway of
Ising model in three-dimensional lattice has also been studied using
transition path sampling approach \cite{JPC0419681}. In addition,
Ising model has been frequently used to test the validity of
classical nucleation theory (CNT)
\cite{EPJ98000571,JCP99006932,JCP00001976, PRE05031601,PRE10030601}.
However, all these studies are limited to regular lattices in
Euclidean space. Since many real systems can be modeled by complex
networks, it is thus natural to ask how the topology of a networked
system would influence the nucleation process of Ising model. To the
best of our knowledge, this has never been studied in the literature
so far.

Although the main motivation of the present study is to address a
fundamental problem in statistical physics, it may also be of
practical interest. For example, our study may help understand how
public opinion or belief changes in the social context
\cite{RMP09000591}, where binary spins can represent two opposite
opinions and the concept of physical temperature corresponds to a
measure of noise due to imperfect information or uncertainty on the
part of the agent. Another example is functional transition in real
biological networks, such as the transition between different
dynamical attractors in neural networks \cite{PNAS04004341}, wherein
the two states of the Ising model may correspond to a neuron being
fired or not. Other examples include gene regulatory networks,
wherein genes can be on or off, corresponding to the two states of
Ising model \cite{PNAS06008372}.

In the present work, we have studied nucleation process of the Ising
model in SFNs. Since nucleation is an activated process that occurs
extremely slow, brute-force simulation is prohibitively expensive.
To overcome this difficulty, we adopt a recently developed forward
flux sampling (FFS) method to obtain the rate and pathway for
nucleation \cite{PRL05018104}. For homogeneous nucleation, we find
that the nucleation begins with nodes with smaller degree, while
nodes with larger degree are more stable. We show that the
nucleation rate decays exponentially  with the network size $N$, and
accordingly the critical nucleus size increases linearly with $N$,
implying that homogeneous nucleation can only occur in finite-size
networks. Comparing the results of networks with different $\gamma$
and those of random networks, we conclude that network heterogeneity
is unfavorable to nucleation. In addition, we have also investigated
heterogeneous nucleation by adding impurities into the networks. It
is found that the dependence of the nucleation rate on the number of
random impurities is significantly different from the case of
targeted impurities. These simulation results may be qualitatively
understood in a mean-field manner.

The rest of the paper is organized as follows. In Sec.\ref{sec2}, we
give the details of our simulation model and the FFS method applied
to this system. In Sec.\ref{sec3}, we present the results for the
nucleation rate and pathway. We then show, via both simulation and
analysis, that the system-size effect of the nucleation rate and
heterogeneous nucleation. At last, discussion and main conclusions
are addressed in Sec.\ref{sec4}.

\section{Model and Simulation details}\label{sec2}

\subsection{The networked Ising model} \label{sec2.1}

The Ising model in a network comprised of $N$ nodes is described by
the Hamiltonian
\begin{equation}
H=-J\sum\nolimits_{i < j}{a_{ij}s_i s_j}-h\sum\limits_i s_i,
\label{eq1}
\end{equation}
where spin variable $s_i$ at node $i$ takes either $+1$ (up) or $-1$
(down). $J$ is the coupling constant and $h$ is the external
magnetic field. The element of the adjacency matrix of the network
takes $a_{ij}=1$ if nodes $i$ and $j$ are connected and $0$
otherwise.

Our simulation is performed by Metropolis spin-flip dynamics
\cite{Lan2000}, in which we attempt to flip each spin once, on
average, during each Monte Carlo (MC) cycle. In each attempt, a
randomly chosen spin is flipped with the probability $\min(1,e^{-
\beta \Delta E})$, where $\beta=1/(k_B T)$ with $k_B$ the Boltzmann
constant and $T$ the temperature, $\Delta E$ is the energy change
due to the flipping process. Generally, with the increment of $T$,
the system will undergo a second-order phase transition at the
critical temperature $T_c$ from an ordered state to a disordered
one. To study nucleation, we set $J=1$, $h>0$, $T<T_c$, and start
from a metastable state in which $s_i=-1$ for most of the spins. The
system will stay in that state for a significantly long time before
undergoing a nucleation transition to the thermodynamic stable state
with most spins pointing up. We are interested in dynamics of this
nucleation process.

\subsection{Forward flux sampling} \label{sec2.2}

FFS method has been used to calculate rate constants, transition
paths and stationary probability distributions for rare events in
equilibrium and nonequilibrium systems \cite{PRL05018104,
PRL06065701, JCP08134704, JPC06004985, JCP07114109, JCP06024102}.
This method uses a series of interfaces in phase space between the
initial and final states to force the system from the initial state
$A$ to the final state $B$ in a ratchet-like manner. An order
parameter $\lambda(x)$ is first defined, where $x$ represents the
phase space coordinates, such that the system is in state $A$ if
$\lambda(x)<\lambda_0$ and state $B$ if $\lambda(x)>\lambda_m$,
while a series of nonintersecting interfaces $\lambda_i$ ($0<i<m$)
lie between states $A$ and $B$, such that any path from $A$ to $B$
must cross each interface without reaching $\lambda_{i+1}$ before
$\lambda_i$. The transition rate $R$ from $A$ to $B$ is calculated
as
\begin{equation}
R=\bar \Phi_{A,0} P\left( {\lambda_{m} |\lambda_0} \right) =\bar
\Phi_{A,0} \prod\nolimits_{i=0}^{m-1}{P\left( {\lambda_{i + 1}
|\lambda_i} \right)}, \label{eq2}
\end{equation}
where $\bar \Phi_{A,0}$ is the average flux of trajectories crossing
$\lambda_0$ in the direction of $B$. $P\left( {\lambda_{m}
|\lambda_0} \right)=\prod\nolimits_{i=0}^{m-1}{P\left( {\lambda_{i +
1} |\lambda_i} \right)}$ is the probability that a trajectory
crossing  $\lambda_0$ in the direction of $B$ will eventually reach
$B$ before returning to $A$, and ${P\left( {\lambda _{i + 1}
|\lambda _i } \right)}$ is the probability that a trajectory which
reaches $\lambda_i$, having come from $A$, will reach
$\lambda_{i+1}$ before returning to $A$. For more information about
FFS, please turn to Ref.\cite{JPH09463102}.

\section{Results}\label{sec3}

\subsection{Homogeneous Nucleation: Rate and pathway} \label{sec3.1}

To begin, we first consider homogeneous nucleation in
Barab\'{a}si--Albert scale-free network (BA-SFN) whose degree
distribution follows a power-law $P(k)\sim k^{-\gamma}$ with the
scaling exponent $\gamma=3$ \cite{SCI99000509}. We define the order
parameter $\lambda$ as the total number of up spins in the networks.
We set $N=1000$, the average degree $\left\langle k \right\rangle
=6$, $T=2.59$, $h=0.7$, $\lambda_0=130$ and $\lambda_m=880$, where
$T$ is lower than the critical temperature $T_c \simeq 10.36$. The
spacing between interfaces is fixed at $3$ up spins, but the
computed results do not depend on this spacing. During the FFS
sampling, we perform $1000$ trials at each interface, from which at
least $100$ configurations are stored in order to investigate the
statistical properties of the ensemble of nucleation pathway. We
obtain $\bar \Phi_{A,0}=1.24 \times 10^{-4} MC step^{-1} spin^{-1} $
and $P\left( {\lambda_{m} |\lambda_0} \right)=4.48 \times 10^{-46}$,
resulting in $R_{Hom}=5.55 \times 10^{-50} MC step^{-1} spin^{-1}$
following Eq.\ref{eq2}. Such a nucleation rate is very low such that
a brute-force simulation would be very expensive.

\begin{figure}
\centerline{\includegraphics*[width=0.85\columnwidth]{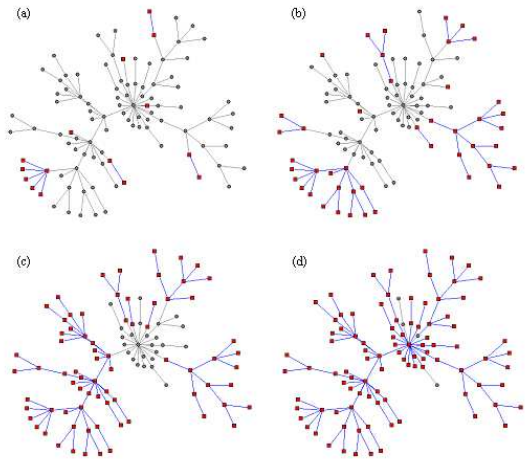}}
\caption{(Color online) Snapshots of nucleation in a BA scale-free
network with $N=100$ and $\left\langle k \right\rangle = 2$ at four
different stages. Up-spins and down-spins are indicated by red
squares and black circles, respectively. \label{fig1}}
\end{figure}

\begin{figure}
\centerline{\includegraphics*[width=0.7\columnwidth]{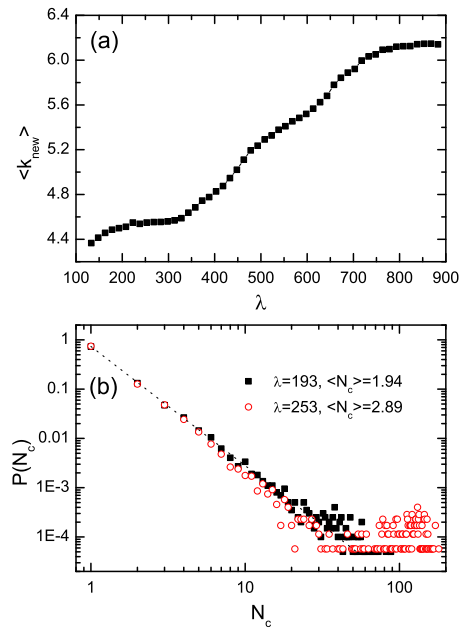}}
\caption{(Color online) (a) The average degree of nodes of new phase
$\left\langle {k_{new} } \right\rangle$ as a function of $\lambda$;
(b) The size distribution of clusters of new phase, in which a power
law distribution is present. Other parameters are $N=1000$,
$\left\langle k \right\rangle =6$, $T=2.59$, and $h=0.7$.
\label{fig2}}
\end{figure}

From the stored configurations at each interface, one can figure out
the details of the nucleation pathway.  Figure \ref{fig1}
illustrates schematically four stages of a typical nucleation
pathway. Clearly, the new phase (indicated by squares) starts from
nodes with smaller degree, while nodes with larger degree are more
stable. This picture is reasonable because nodes with larger degrees
need to overcome more interfacial energies. Figure \ref{fig2}(a)
plots the average degree of network nodes in the new phase, $
\left\langle {k_{new}} \right\rangle $, as a function of the order
parameter $\lambda$. As expected, $\left\langle {k_{new} }
\right\rangle $ increases monotonously with $\lambda$. On the other
hand, it is observed that the formation of large clusters of new
phase is accompanied with the growth and coalescence of small
clusters. Interestingly, we find that the size $N_c$ of new phase
clusters follows a power law distribution at early stages of
nucleation, $P(N_c) \sim N_c^{-\alpha}$ with the fitting exponent
$\alpha\simeq2.44$, as shown in Fig.\ref{fig2}(b). With the
emergence of a giant component of new phase, the tail of the
distribution is elevated, but the size distribution for the
remaining clusters still follows power-law. The underlying mechanism
of such phenomenon is still an open question for us.

\begin{figure}
\centerline{
\includegraphics*[width=0.9\columnwidth]{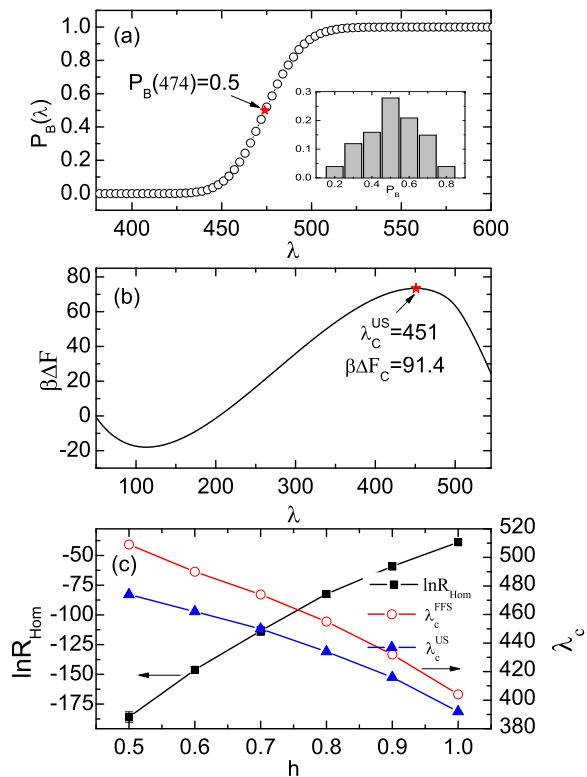}}
\caption{(Color online) (a) The committor probability $P_B$ as a
function of $\lambda$; The inset plots the committor distribution at
$\lambda_c^{FFS}$. (b) The free energy $\Delta F$ as a function of
$\lambda$, in which the maximum in $\Delta F$ occurs at
$\lambda_c^{US}$. (c) The logarithm of homogeneous nucleation rate
$\ln R_{Hom}$ (left axis), and the critical size of nucleus,
$\lambda_c^{FFS}$ and $\lambda_c^{US}$ (right axis), obtained by FFS
method and US method, respectively, as functions of $h$. Other
parameters are the same as Fig.\ref{fig2}. \label{fig3}}
\end{figure}

To determine the critical size of the nucleus, $\lambda_c$, we
compute the committor probability $P_B$, which is the probability of
reaching the thermodynamic stable state before returning to the
metastable state.  The dependence of $P_B$ on $\lambda$ is plotted
in Fig.\ref{fig3}(a). As commonly reported in the literature
\cite{JPC0419681, PRE10030601}, the critical nucleus appears at
$P_B(\lambda_c)=0.5$, giving the critical nucleus size
$\lambda_c^{FFS}=474$. The committor distribution at
$\lambda_c^{FFS}$ exhibits a peak at $0.5$, of which $70 \%$ of spin
configurations have $P_B$ values within the range of $0.4$ to $0.6$
(see the inset of Fig.\ref{fig3}(a)), indicating that $\lambda$ is a
proper order parameter.

Note that conventionally the nucleation threshold $\lambda_c$ is
usually estimated by using CNT
\cite{AP35000719,AP43000001,ARPC95000489}. One can calculate the
free energy change along the nucleation path, $\Delta F(\lambda)$,
by using methods like umbrella sampling (US) \cite{JCP92000015}.
According to CNT, $\Delta F$ will bypass a maximum at
$\lambda=\lambda_c^{US}$, and the nucleation rate is given by $\nu
\exp ( - \beta \Delta F_c)$, where $\nu$ is an attempt frequency.
Here we have computed $\Delta F$ by using US, in which we have
adopted a bias potential $0.1k_B T(\lambda-\bar \lambda)^2$, with
$\bar \lambda$ being the center of each window. As shown in
Fig.\ref{fig3}(b), the maximum in $\Delta F$ occurs at
$\lambda_c^{US}=451$, giving a free-energy barrier of $\Delta F_c
\simeq 91.4 k_B T$. Clearly $\lambda_c^{US}$ gives a fairly good
estimation of $\lambda_c^{FFS}$. To calculate the nucleation rate,
however, one has to obtain the attempt frequency $\nu$, which is not
a trivial task. If we just set $\nu=1$, we obtain a CNT prediction
of a rate of $2.02 \times 10^{-40} MC step^{-1} spin^{-1}$, which is
$9$ orders of magnitude faster than that computed from FFS method.
This level of disagreement in nucleation rate corresponds to an
error in the free-energy barrier of about $24\%$. Since the accurate
value of $\nu$ is generally unavailable, we will use FFS method to
calculate the nucleation rate throughout this paper. In addition,
real nucleation pathway cannot be obtained by conventional US method
due to the use of a biased potential.

We have also investigated how the nucleation rate and threshold
depend on the external field $h$. In Fig.\ref{fig3}(c), $\ln
R_{Hom}$, $\lambda_c^{FFS}$ and $\lambda_c^{US}$ are plotted as
functions of $h$. The error-bars are obtained via $20$ different
network realizations and $10$ independent FFS samplings. As
expected, $\ln R_{Hom}$ increases monotonously with $h$, and
$\lambda_c^{FFS}$ and $\lambda_c^{US}$ both decrease with $h$. For
large $h$, the difference between $\lambda_c^{FFS}$ and
$\lambda_c^{US}$ becomes small. If $h$ is large enough, one expects
that the free-energy barrier will disappear, and nucleation will be
not relevant.

\subsection{Homogeneous Nucleation: System-size effects} \label{sec3.2}

According to Fig.\ref{fig3}, one finds that nearly half of the nodes
must be inverted to achieve nucleation. This means that for a large
network, nucleation is very difficult to occur. An interesting
question thus arises: How the nucleation rate and threshold depend
on the network size?

To answer this question, we have performed extensive simulations to
calculate $R_{Hom}$ and $\lambda_c$ for different network size $N$.
In particular, besides the BA-SFNs, we have also considered
different network types, including SFNs with other scaling exponent
$\gamma$ and homogeneous random networks (HoRNs) \cite{PRE05056128}.
The networks are generated according to the Molloy-Reed model
\cite{RSA95000161}: Each node is assigned a random number of stubs
$k$ that are drawn from a specified degree distribution. This
construction eliminates the degree correlations between neighboring
nodes. We note here that the exponent $\gamma$ can be a measure of
degree heterogeneity of the network, i.e., the smaller $\gamma$ is,
the degree distribution is more heterogeneous. In a HoRN, each node
is equivalently connected to other $\left\langle k \right\rangle$
nodes, randomly selected from the whole network, such that no degree
heterogeneity exists. By comparing the results in SFNs with
different $\gamma$ as well as that in HoRNs, one can on one hand,
check the robustness of the system size effects, and on the other
hand, investigate how the degree heterogeneity affects the
nucleation process.

\begin{figure}
\centerline{\includegraphics*[width=0.8\columnwidth]{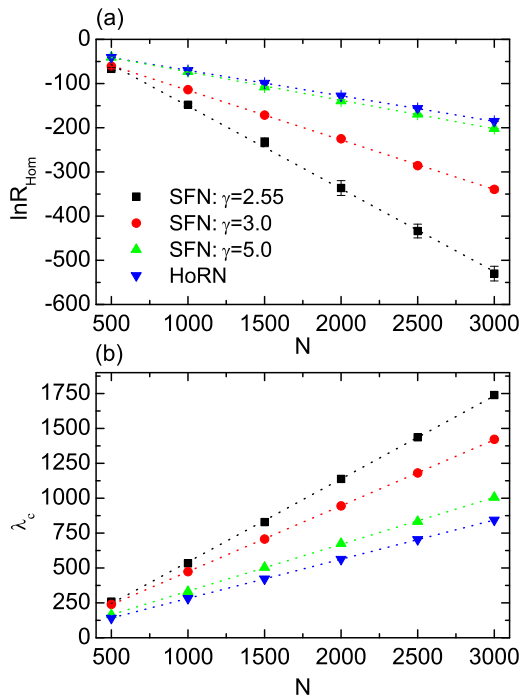}} 
\caption{ (Color online) (a) The logarithm of homogeneous nucleation
rate $\ln R_{Hom}$ and (b) the critical size of nucleus $\lambda_c$
as functions of the network size $N$ in SFNs with different $\gamma$
and in HoRNs. Other parameters are the same as Fig.\ref{fig2}.
\label{fig4}}
\end{figure}

Figure \ref{fig4} shows the simulation results. All the parameters
are the same as in Fig.2, expect that $N$ varies from $N=500$ to
$N=3000$. Interestingly, both $\ln R_{Hom}$ and $\lambda_c$ show
very good linear dependences on the system size, i.e., $\ln R_{Hom}
\sim - aN$ and $\lambda_c \sim bN$ with $a$ and $b$ being positive
constants. Obviously, in the thermodynamic limit $N \rightarrow
\infty$, we have $R_{Hom} \rightarrow 0$ and $\lambda_c \rightarrow
\infty$. This means that nucleation in these systems is not relevant
in the thermodynamic limit, and only finite-size systems are of
interest. As shown in Fig.4, for the network types considered here,
qualitative behaviors are the same. Quantitatively, with increasing
$\gamma$, the line slope becomes smaller, $R_{Hom}$ becomes larger
and $\lambda_c$ gets smaller. Since larger $\gamma$ corresponds to
more homogeneous degree distribution, these results indicate that
the degree heterogeneity is unfavorable to nucleation. This is
consistent with the nucleation pathway as shown in Fig.2: In a
heterogenous network, those hub nodes are difficult to flip, making
the nucleation difficult.

In the following, we will show that the system-size effects can be
qualitatively understood by CNT and simple mean-field (MF) analysis.
According to CNT, the formation of a nucleus lies in two competing
factors: the energy cost of creating a new up spin which favors the
growth of the nucleus, and an opposing factor which is due to the
creation of new interfaces between up and down spins. The change in
the free energy may be written as
\cite{AP35000719,AP43000001,ARPC95000489}
\begin{equation}
\Delta F(\lambda) =-2h \lambda + \sigma \lambda, \label{eq3}
\end{equation}
where $\sigma$ denotes the effective interfacial free energy, which
may depend on $T$, $h$, and $N$. Since interfacial interactions
arise from up spins inside the nucleus and down spins outside it,
one may write $\sigma=2J K_{out}$ by neglecting entropy effects
(zero-temperature approximation), where $K_{out}$ is the average
number of neighboring down-spin nodes that an up-spin node has.
Using MF approximation, one has $K_{out}=\left\langle k
\right\rangle(1 - \lambda/N)$. Inserting this relation to
Eq.\ref{eq3} and maximizing $\Delta F$ with respect to $\lambda$, we
have
\begin{equation}
\lambda_c^{MF}= {{J\left\langle k \right\rangle - h} \over
{2J\left\langle k \right\rangle}}N, \label{eq4}
\end{equation}
and the free-energy barrier
\begin{equation}
\Delta F_c^{MF} = {{(J\left\langle k \right\rangle - h)^2 N} \over
{2J\left\langle k \right\rangle}}. \label{eq5}
\end{equation}
Clearly, both $\lambda_c^{MF}$ and $\Delta F_c^{MF}$ linearly
increase with $N$ if other parameters are fixed. Therefore, the
linear relationships shown in Fig.\ref{fig4} are essentially
analogous to the behavior of a mean-field network. Quantitatively,
however, the MF analysis fails to predict the line slopes in
Fig.\ref{fig4}. This can be understood because the approximations
are so crude, wherein important aspects such as network
heterogeneity and entropy effects have not been accounted for. A
rigid analysis is not a trivial task and beyond the scope of the
present work.

\subsection{Heterogeneous nucleation} \label{sec3.3}

\begin{figure}
\centerline{\includegraphics*[width=0.8\columnwidth]{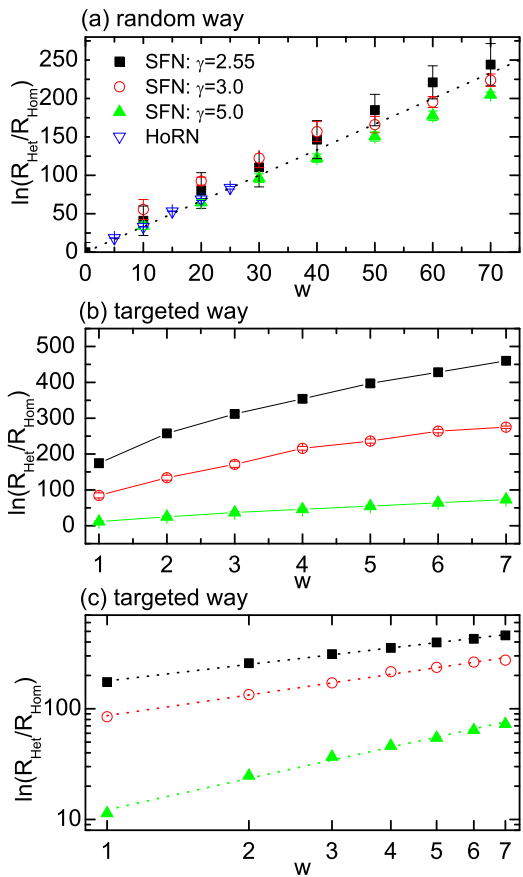}}
\caption{ (Color online) $\ln \left( {\frac{R_{Het}} {{R_{Hom} }}}
\right)$ as a function of the number of impurity nodes $w$. (a)
corresponds to the case of random impurities, while (b) and (c) to
the case of targeted impurities. (c) plots $\ln \left(
{\frac{R_{Het}} {{R_{Hom} }}} \right)$ \emph{vs} $w$ in double
logarithmic coordinates. All dotted lines are drawn by linear
fitting. Other parameters are the same as Fig.\ref{fig2} except for
$h=0.2$. \label{fig5}}
\end{figure}

In practice, most nucleation events that occur in nature are
heterogeneous, i.e., impurities of the new phase are initially
present. It is well known that impurities can increase nucleation
rate by as much as several orders of magnitude. In our model,
impurities are introduced by fixing some nodes in up-spin state. We
are interested in how the number $w$ of impurity nodes and the way
of adding impurities would affect the nucleation rate. The first way
of adding impurities we use is that impurity nodes are selected in a
random fashion. Figure \ref{fig5}(a) gives the simulation results of
$\ln \left( {\frac{R_{Het}} {{R_{Hom} }}} \right)$ as a function of
$w$ in different network models, where $R_{Het}$ are the rates of
heterogeneous nucleation. As expected, nucleation becomes faster in
the presence of random impurities no matter which kind of network
model is applied. It seems that in Fig.\ref{fig5}(a) all data
collapse and exhibit a linear dependence on $w$, with the fitting
slope $3.33$. This means that each additional random impurity can
lead to the increase of the rate by more than one order of
magnitude. For the second way, we select $w$ nodes with most highly
degree as the impurity nodes, termed as targeted impurities.
Strikingly, such a targeted scheme is much more effective in
increasing the nucleation rate than random one, as shown in
Fig.\ref{fig5}(b). For example, for SFNs with $\gamma=3$, one single
targeted impurity can increase the rate by about $36$ orders of
magnitude.

As in Sec.\ref{sec3.2}, below we will also give a MF analysis of the
heterogeneous nucleation, which qualitatively agrees with the
simulation results. Each impurity node contributes an additional
term to the free energy barrier, which can, under zero-temperature
approximation, be written as the product of $-2J$ and the expected
degree of the impurity node. For random impurities, each impurity
node has an expected degree $\left\langle k \right\rangle$, yielding
the term $-2J\left\langle k \right\rangle$. Thus, the resulting
free-energy barrier of the heterogenous nucleation becomes $\Delta_1
F^{Het}_c = \Delta F^{Hom}_c - 2J\left\langle k \right\rangle w,$
where $\Delta F^{Hom}_c$ is the free-energy barrier of homogeneous
nucleation. According to CNT, one obtains
\begin{equation}
\ln \left( {\frac{R_{Het}} {{R_{Hom} }}} \right) =
\frac{{2J\left\langle k \right\rangle }} {{k_B T}}w. \label{eq6}
\end{equation}
Therefore, nucleation with random impurities is always faster than
without impurity, and $\ln \left( {\frac{R_{Het}} {{R_{Hom} }}}
\right)$ should vary linearly with $w$. The theoretical estimate of
the slope is given by $2J\left\langle k \right\rangle/k_BT=4.63$,
approximately consistent with the simulation one. Given the simple
nature of the above approximation the agreement is satisfactory. For
the targeted way, a similar treatment to the former case can also be
executed, except that $\left\langle k \right\rangle$ should be
replaced by $\left\langle k \right\rangle_w$, where $\left\langle k
\right\rangle_w$ is the average degree of the $w$ targeted nodes.
After simple calculations, we can obtain $ \left\langle k
\right\rangle_w=\left\langle k \right\rangle (\frac{N}
{w})^{\frac{1} {{\gamma-1}}}$. This leads to a free-energy barrier,
$\Delta _2 F_c^{Het} = \Delta _2 F_c^{Hom}  - 2J\left\langle k
\right\rangle N^{\frac{1} {{\gamma  - 1}}} w^{\frac{{\gamma-2}}
{{\gamma-1}}}$, and
\begin{equation}
\ln \left( {\frac{R_{Het}} {{R_{Hom} }}} \right) =
\frac{{2J\left\langle k \right\rangle }} {{k_B T}}N^{\frac{1}
{{\gamma  - 1}}} w^{\frac{{\gamma  - 2}} {{\gamma  - 1}}}.
\label{eq7}
\end{equation}
Compared with Eq.\ref{eq6}, besides the presence of an additional
size-dependent factor of $N^{\frac{1} {{\gamma  - 1}}}$, the
$w$-dependent factor becomes $w^{{\frac{{\gamma  - 2}} {{\gamma -
1}}}}$ rather than $w$. For homogeneous network, i.e. $\gamma \to
\infty$, one obtains $N^{\frac{1}{{\gamma - 1}}} \to 1$ and
$w^{\frac{{\gamma - 2}} {{\gamma - 1}}} \to w$, and Eq.\ref{eq7} is
thus equivalent to Eq.\ref{eq6}. In Fig.\ref{fig5}(c), we plot the
simulation results of $\ln \left( {\frac{R_{Het}} {{R_{Hom} }}}
\right)$ as a function of $w$ in double logarithmic coordinates,
where linear dependences are apparent, in agreement with the result
of Eq.\ref{eq7}. The fitting slopes $\mu$ and intercepts $\kappa$
for $\gamma=2.55, 3.0, 5.0$ in Fig.\ref{fig5}(c) are $\mu=0.49,
0.62, 0.93$, and $\kappa=2.25, 1.94, 1.08$, respectively, while our
analytical estimates are $\mu=\frac{{\gamma-2}}{{\gamma-1}}=0.35,
0.50, 0.75$, and $\kappa=\log \left( {\frac{{2J\left\langle k
\right\rangle N^{\frac{1} {{\gamma-1}}} }} {{k_B T}}} \right)=2.60,
2.16, 1.42$, respectively. Simulations and analysis give the same
trends of the values of $\mu$, $\kappa$ and $\gamma$.

\section{Discussion and Conclusions} \label{sec4}

Our investigations of system-size effects of nucleation in SFNs and
HoRNs have revealed that the nucleation rate is size-dependent, and
it decreases exponentially with the network size, resulting in that
nucleation only occurs at finite-size systems. As we already know,
such system-size dependence does not exist in two-dimensional
regular lattices \cite{JPC06004985}. We have also studied nucleation
of Ising system in regular networks where each node is connected to
its $k$-nearest neighbors (here we only consider the case of sparse
networks, i.e., $k \ll N$), and found that the rate is almost
independent of network size (results not shown). Therefore,
nucleation process in SFNs and HoRNs are quite different from that
in regular lattices or networks. Such differences may originate from
the infinite-dimensional properties of SFNs and HoRNs, wherein the
average path distance is rather small, rendering the system's
behavior analogous to that of a mean-field network. An interesting
situation arises when one considers Watts-Strogatz small-world
network, which is constructed by randomly rewiring each link of a
regular network with the probability $p$ \cite{NAT98000440}. With
the increment of $p$ from $0$ to $1$, the resulting network changes
from a regular network to a completely random one. As mentioned
above, for the nucleation process, no system-size effects exist for
$p=0$, while system-size dependence exists for $p=1$. One may
naturally ask: How does the crossover happens when $p$ changes from
$0$ to $1$, and what is the physical significance of such a
transition? This question surely deserves further investigations and
may be the content of a future publication.

In summary, we have studied homogeneous and heterogeneous nucleation
of Ising model in SFNs, by using FFS method. For homogeneous
nucleation, we find that the formation of new phase starts from
nodes with smaller degree, while nodes with higher degree are more
stable. Extensive simulations show that the nucleation rate
decreases exponentially with the network size $N$, and the
nucleation threshold increases linearly with $N$, indicating that
nucleation in these systems are not relevant in the thermodynamic
limit. For heterogeneous nucleation, target impurities are shown to
be much more efficient to enhance the nucleation rate than random
ones. Simple MF analysis is also present to qualitatively illustrate
the simulation results. Our study may provide valuable understanding
how first-order phase transition takes place in network-organized
systems, and how to effectively control the rate of such a process.

\begin{acknowledgments}
This work is supported by the National Science Foundation of China
under Grants No. 20933006 and No. 20873130.
\end{acknowledgments}

%

\end{document}